\documentstyle[prb,aps,floats,epsfig,eqsecnum]{revtex}
\preprint{ITP-SB-96-58}
\begin{document}
\twocolumn[\hsize\textwidth\columnwidth\hsize\csname@twocolumnfalse
\endcsname
\title{Exact two$-$spinon dynamic structure factor \\
  of the one$-$dimensional $s=1/2$ Heisenberg$-$Ising antiferromagnet} 
\author{A. Hamid Bougourzi} 
\address{Institute of Theoretical Physics, SUNY at Stony
  Brook, Stony Brook, NY 11794} 
\author{Michael Karbach\cite{AAAuth} and Gerhard M\"uller}
\address{Department of Physics, The University of Rhode Island,
Kingston RI 02881-0817}

\date{\today}
\maketitle
%
%
\begin{abstract}
%
%
  The exact 2-spinon part of the dynamic spin structure factor
  $S_{xx}(Q,\omega)$ for the one-dimensional $s$=1/2 $XXZ$ model at $T$=0 in the
  antiferromagnetically ordered phase is calculated using recent
  advances by Jimbo and Miwa in the algebraic analysis based on
  (infinite-dimensional) quantum group symmetries of this model and the related
  vertex models. The 2-spinon excitations form a 2-parameter continuum
  consisting of two partly overlapping sheets in $(Q,\omega)$-space. The
  spectral threshold has a smooth maximum at the Brillouin zone boundary
  $(Q=\pi/2)$ and a smooth minimum with a gap at the zone center $(Q=0)$. The
  2-spinon density of states has square-root divergences at the lower and upper
  continuum boundaries. For the 2-spinon transition rates, the two regimes $0
  \leq Q < Q_\kappa$ (near the zone center) and $Q_\kappa < Q \leq \pi/2$ (near
  the zone boundary) must be distinguished, where $Q_\kappa \rightarrow 0$ in
  the Heisenberg limit and $Q_\kappa \rightarrow \pi/2$ in the Ising limit. In
  the regime $Q_\kappa < Q \leq \pi/2$, the 2-spinon transition rates relevant
  for $S_{xx}(Q,\omega)$ are finite at the lower boundary of each sheet,
  decrease monotonically with increasing $\omega$, and approach zero linearly at
  the upper boundary. The resulting 2-spinon part of $S_{xx}(Q,\omega)$ is then
  square-root divergent at the spectral threshold and vanishes in a square-root
  cusp at the upper boundary. In the regime $0 < Q_\kappa \leq \pi/2$, by
  contrast, the 2-spinon transition rates have a smooth maximum inside the
  continuum and vanish linearly at either boundary. In the associated 2-spinon
  line shapes of $S_{xx}(Q,\omega)$, the linear cusps at the continuum
  boundaries are replaced by square-root cusps. Existing perturbation studies
  have been unable to capture the physics of the regime $Q_\kappa < Q \leq
  \pi/2$. However, their line shape predictions for the regime $0 \leq Q <
  Q_\kappa$ are in good agreement with the new exact results if the anisotropy
  is very strong. For weak anisotropies, the exact line shapes are more
  asymmetric.

\end{abstract}
\pacs{??}
\twocolumn
]
%
%
\section{Introduction}\label{secI}
%
%
Among all the spin-chain models that are directly relevant for the description
of real quasi-one-dimensional (1D) magnetic insulators, the $s=1/2$ $XXZ$ model,
\begin{equation}\label{H}
  H = -\frac{J}{4} \sum_{n=-\infty}^{\infty} 
  (\sigma_n^x \sigma_{n+1}^x +\sigma_n^y\sigma_{n+1}^y + 
  \Delta\sigma_n^z \sigma_{n+1}^z), 
\end{equation}
is the one whose physical properties have been studied most comprehensively.
Today there exist more exact results for this model than for any other model of
comparable importance.

The early demonstration\cite{Orba58,dCG66,YY66a} that the $XXZ$ model is
amenable to the Bethe ansatz led to a steady stream of advances in our
understanding of many of its ground-state properties,\cite{LP75b,KM93} its
thermodynamic properties,\cite{Gaud71,JM72,John74} and the structure of its
excitation spectrum.\cite{HB74,TF79,Mull82,MM83,IS80,MTBB81,FT81}
The $T=0$ phase diagram of the $XXZ$ model, which was rigorously established by
these advances, consists of a ferromagnetic phase at $\Delta\geq 1$, a critical
phase (spin-fluid, Luttinger liquid) at $-1 \leq \Delta < 1 $, and an
antiferromagnetic phase at $\Delta < -1$.

The mapping between the $XXZ$ model and the exactly solvable 6-vertex and
8-vertex models yielded additional ground-state properties of the former on a
rigorous basis, notably the spontaneous staggered magnetization in the
antiferromagnetic phase and some critical exponents in the spin-fluid
phase.\cite{JKM73,LP75b,Baxt82}

Until recently, exact results for the $T=0$ spin dynamics of the $XXZ$ model
were limited to a single nontrivial case, the $XX$ model ($\Delta=0$). For this
case, the spin system is equivalent to a system of free lattice
fermions,\cite{LSM61} and the dynamic spin correlation functions can be
expressed as fermion density correlations ($zz$)\cite{Niem67} or as infinite
determinants or Pfaffians ($xx$).\cite{MBA71,LP75a} In the surrounding
spin-fluid phase $(-1 \leq \Delta < 1)$, exact results for the infrared
singularities of dynamic structure factors were obtained by field-theoretic
approaches.\cite{LP75b,Foge78}

A new avenue for the study of the $T=0$ dynamics of the $XXZ$ model on a
rigorous basis was opened up by important advances in the study of this model
and the related vertex models in the framework of the algebraic analysis based
on quantum group symmetries. A detailed description of this method with all the
results which our calculations build on can be found in a recent book by Jimbo
and Miwa.\cite{JM95} Unlike the Bethe ansatz, this approach considers an {\it
  infinite} chain from the outset and exploits the higher symmetry of the
infinite system (compared to the finite system) described by the quantum group
$U_q(sl_2)$.\cite{note5,GRS96}

The algebraic analysis of the $XXZ$ model for the purpose of calculating
correlation functions and transition rates (form factors) of local spin
operators requires the execution of the following program: (i) span
the infinite-dimensional physically relevant Hilbert space in the form of a
separable Fock space of multiple spinon excitations and to generate the $XXZ$
eigenvectors in this Fock space by products of spinon creation operators
(so-called vertex operators) from the $XXZ$ ground state (physical vacuum); (ii)
determine the spectral properties (energy, momentum) of the spinon
excitations; (iii) express the local spin operators in terms of vertex
operators; and (iv) evaluate matrix elements of products of vertex operators
in this spinon eigenbasis.

There exist two similar yet distinct programs which operate under different
circumstances for essentially the same purpose. One is the fermion
representation of the 1D $s=1/2$ $XY$ model or the equivalent 2D Ising
model,\cite{MBA71,WMTB76} and the other is conformal field theory for critical
(massless) continuum models.\cite{BPZ84} Quantum inverse scattering theory
provides yet different ways of calculating some correlation functions and matrix
elements for massive relativistic continuum models\cite{Smir92} and for the
$XXX$ model.\cite{KBI93}

The algebraic analysis\cite{JM95} operates in the massive phase stabilized by
N\'eel long-range order at $\Delta<-1$, but the isotropic limit
$\Delta \to -1^-$ can be performed meaningfully at various stages of the
calculation and thus yields equivalent results for the (massless) Heisenberg
antiferromagnet.\cite{BCK96,KMB+97}

In this paper we infer from the diverse ingredients now accessible via the Bethe
ansatz and the algebraic analysis, an explicit expression for the exact 2-spinon
part of the dynamic spin structure factor
\begin{equation}\label{stfdef}
  S_{xx}(Q,\omega)=\frac{1}{4}\sum_{n=-\infty}^{+\infty}
  \int\limits_{-\infty}^{\infty} dt\; e^{i(\omega t+Qn)} 
  \langle \sigma_n^x(t)\sigma_0^x\rangle,
\end{equation}
at $T=0$ and $\Delta<-1$. The line shapes thus obtained are of direct relevance
for the interpretation of existing spectroscopic data obtained via inelastic
neutron scattering\cite{YHSS81,NBAB83a,GTN95} and Raman scattering\cite{LBW81}
on the quasi-1D magnetic compounds $\rm CsCoCl_3$ and $\rm CsCoBr_3$.

In Sec.~\ref{secII} we discuss the $m$-spinon eigenbasis and infer a suitable
parametric representation of the energy-momentum relation for spinon excitations
from it. In Sec.~\ref{secIII} a closed-form expression for the
2-spinon density of states $D(Q,\omega)$ is derived from this spectral
information. In Sec.~\ref{secIV} we analyze the matrix elements (form factors)
between the twofold degenerate ground state and the 2-spinon excitations and
derive from them (in Sec.~\ref{secVI}), after having solved the 2-spinon
energy-momentum relations in the appropriate parametrization (Sec.~\ref{secV}),
a function $M(Q,\omega)$ which, when multiplied with $D(Q,\omega)$, yields the
2-spinon dynamic structure factor $S_{xx}^{(2)}(Q,\omega)$.

A related study was previously undertaken by Weston and Bougourzi.\cite{WB94} In
that study the goal was to calculate the 2-spinon part of the dynamic structure
factor $S(Q,\omega)$ defined as the Fourier transform of
$\langle\mbox{\boldmath$\sigma$}_n(t)\cdot\mbox{\boldmath$\sigma$}_0\rangle$.
The result was expressed as an expansion about the Ising limit $(\Delta\to
-\infty)$ carried out explicitly to $12^{\rm th}$ order. It is much more
difficult to calculate this quantity than to calculate $S_{xx}^{(2)}(Q,\omega)$,
where no expansion is necessary to obtain explicit results.

Finally, it is interesting to note that the exact result for the
frequency-dependent spin autocorrelation function $\Phi_{xx}(\omega)\equiv
\int_{-\pi}^{\pi}(dQ/2\pi)S_{xx}(Q,\omega)$ of the case $\Delta=0$, which was
calculated in the fermion representation,\cite{MS84a} represents all $m$-spinon
contributions for $m=2,4,\ldots$ simultaneously. There the $m$-spinon structure
of the excitation spectrum is reflected in $\Phi_{xx}(\omega)$ by an infinite
sequence of singularities at the band-edge frequencies $\omega/J=0,1,2,\ldots$.
%
%
\section{Spectrum}\label{secII}
%
%
The $2^N$-dimensional Hilbert space of the $XXZ$ model for a chain of $N$ sites
becomes non-separable in the limit $N\to\infty$. However, for the infinite
chain, a separable subspace $\cal F$ can be constructed, and all physical
properties of the $XXZ$ model can, in principle, be derived exactly from it. The
classification of the $XXZ$ spectrum in terms of $m$-spinon excitations, which
is instrumental in the quantum group analysis, had already been established by
Faddeev and Takhtajan,\cite{TF79,FT81} for $\Delta=-1$ in the framework of
the algebraic Bethe ansatz.

The (infinite-dimensional) space $\cal F$ is spanned by vectors
$|\xi_m,\epsilon_m;\ldots;\xi_1,\epsilon_1\rangle_j$ with $m=0,1,\ldots$ and
$j=0,1$, which represent multiple spinon excitations.  In the regime of interest
here, the twofold degenerate vacuum state is represented by the two vectors
$|0\rangle_0, |0\rangle_1$. These states break the translational symmetry of
$H$. The translation operator $T$ (shift by one lattice site) transforms the two
vectors into each other:
\begin{equation}
  \label{transdef}
  T|0\rangle_j = |0\rangle_{1-j}, \quad j=0,1.
\end{equation}
In the Ising limit ($\Delta\to -\infty$), they become the pure N\'eel states,
$|\ldots\uparrow\downarrow\uparrow\downarrow\ldots\rangle,
|\ldots\downarrow\uparrow\downarrow\uparrow\ldots\rangle$.

Each spinon excitation is characterized by a (complex) spectral parameter
$\xi_l$ and a spin orientation $\epsilon_l=\pm 1$. The subspaces of $\cal F$
with even and odd numbers of spinon excitations are disconnected in
all matters of concern here. They describe the physics of chains with even and
odd $N$ asymptotically for $N\to\infty$.\cite{TF79,FT81} The
completeness relation for the spinon basis in $\cal F$ reads\cite{JM95}
\begin{eqnarray}\label{comrel} 
    {\bf I}&=&\sum_{j=0,1}
    \sum_{m=0,1,\ldots}^\infty 
    \sum_{\epsilon_1,\ldots,\epsilon_m=\pm 1}\frac{1}{m!}
    \oint \prod_{i=1}^m \frac{d\xi_i}{2\pi i \xi_i} 
\nonumber \\ && \times
    |\xi_m,\epsilon_m;\ldots;\xi_1,\epsilon_1\rangle_j
    {_j\langle\xi_1,\epsilon_1;\ldots;\xi_m,\epsilon_m|}. 
\end{eqnarray}
These basis vectors are, in fact, eigenvectors of the $XXZ$ Hamiltonian $H$ and
of the translation operator $T^2$,
\begin{mathletters}\label{states}
\begin{eqnarray} 
    T|\xi_m\!,\!\epsilon_m\!;\!\ldots\!;\!\xi_1\!,\!\epsilon_1\rangle_j 
&=&
    \prod_{i=1}^m\frac{1}{\tau(\xi_i)}
    |\xi_m\!,\!\epsilon_m\!;\!\ldots\!;\!\xi_1\!,\!\epsilon_1\rangle_{1\!-\!j}, 
\\ 
    H|\xi_m\!,\!\epsilon_m\!;\!\ldots\!;\!\xi_1\!,\!\epsilon_1\rangle_j 
&=&
    \sum_{i=1}^m e(\xi_i)
    |\xi_m\!,\!\epsilon_m\!;\!\ldots\!;\!\xi_1\!,\!\epsilon_1\rangle_j,
\end{eqnarray}  
\end{mathletters}
with the respective eigenvalues determined by 
\begin{mathletters}\label{enmom}
\begin{eqnarray}\label{tauxi}
    \tau(\xi)
&=& e^{-ip(\xi)} = 
    \xi^{-1} \frac{\theta_{q^4}(q\xi^2)}{\theta_{q^4}(q \xi^{-2})}, \\
    e(\xi) 
&=&
    J\frac{1-q^2}{4q} \xi \frac{d}{d\xi} \log \tau(\xi),
\end{eqnarray}
\end{mathletters}
in terms of the spectral parameter $\xi$ and the anisotropy parameter 
\begin{equation}
  \label{Delta}
    \Delta=(q+q^{-1})/2,~~-1<q<0.
\end{equation}
Here $q$ is the deformation parameter of the quantum group
$U_q(sl_2)$,\cite{note5} and
\begin{mathletters}\label{jmdef}
\begin{eqnarray}
    \theta_x(y)       &\equiv& \biglb(x;x\bigrb)\biglb(y;x\bigrb)
                       \biglb(x y^{-1};x\bigrb), \\
    \biglb(y;x\bigrb) &\equiv& \prod_{n=0}^{\infty} (1-y x^n).
\end{eqnarray}
\end{mathletters}

For most of the analysis to be carried out later, it is convenient to express
$\xi$ in terms of the alternative spectral parameter $\beta$:
\begin{equation}\label{alphabeta}
    \xi \equiv i e^{i\pi\beta/2K},~~ -2K \leq \beta < 2K.
\end{equation}
The energy and momentum of a spinon are then expressed in terms of Jacobian
elliptic functions,
\begin{mathletters}\label{pealpha}
\begin{eqnarray}
  \label{ealpha}
  e(\xi)=\bar e(\beta) 
&=&
  I {\rm dn}\, \beta, \\
  \label{palpha} p(\xi) = \bar p(\beta)
&=&
  {\rm am}\, \beta + \frac{\pi}{2},
\end{eqnarray}
\end{mathletters}
with 
\begin{equation}\label{ampli}
  I\equiv\frac{JK}{\pi}\sinh\frac{\pi K'}{K}.
\end{equation}
The anisotropy parameter (\ref{Delta}) is related to the nome
\begin{equation}\label{nome}
    -q = \exp(-\pi K^\prime/K), 
\end{equation}
and thus determines the moduli $k,k'\equiv \sqrt{1-k^2}$ of the elliptic
integrals $K\equiv K(k), K'\equiv K(k')$.  The spinon energy-momentum relation
resulting from (\ref{pealpha}),
\begin{equation}\label{e1p}
e_1(p) = I\sqrt{1-k^2\cos^2p},\;\; 0\leq p\leq\pi,
\end{equation}
is equivalent to the corresponding relation obtained via Bethe
ansatz.\cite{JM72,FT81,BdVV83}

For the calculation of $S_{xx}^{(2)}(Q,\omega)$ from the 2-spinon density of
states and the 2-spinon matrix elements we introduce here translationally
invariant vacuum states,
\begin{equation}
  \label{tev0pi}
  |0\rangle \equiv \frac{|0\rangle_0 + |0\rangle_1}{\sqrt{2}}, 
  \qquad
  |\pi\rangle \equiv \frac{|0\rangle_0 - |0\rangle_1}{\sqrt{2}}, 
\end{equation}
which have wave numbers (total momenta mod $2\pi$) $0$ and $\pi$, respectively,
in the extended Brillouin zone $(-\pi,+\pi)$.\cite{note2} In the isotropic
limit, the state $|0\rangle$ is a singlet $(S_T=0)$, and the state $|\pi\rangle$
is the vector with $S_T^z=0$ of a triplet $(S_T=1)$. The corresponding linear
combinations of 2-spinon states,
\begin{eqnarray}
  \label{tev0pi2m}
  |\xi_2,\epsilon_2;\xi_1,\epsilon_1;0\rangle &\equiv& 
    \frac{|\xi_2,\epsilon_2;\xi_1,\epsilon_1\rangle_0 + 
                            |\xi_2,\epsilon_2;\xi_1,\epsilon_1\rangle_1}
                        {\sqrt{2}}, \nonumber \\
  |\xi_2,\epsilon_2;\xi_1,\epsilon_1;\pi\rangle &\equiv& 
    \frac{|\xi_2,\epsilon_2;\xi_1,\epsilon_1\rangle_0 - 
                            |\xi_2,\epsilon_2;\xi_1,\epsilon_1\rangle_1}
                        {\sqrt{2}},
\end{eqnarray}
are then also translationally invariant,
\begin{eqnarray}
  T|\xi_2,\epsilon_2;\xi_1,\epsilon_1;0\rangle &=&
  e^{i[p(\xi_1)+p(\xi_2)]}|\xi_2,\epsilon_2;\xi_1,\epsilon_1;0\rangle, \\
  T|\xi_2,\epsilon_2;\xi_1,\epsilon_1;\pi\rangle &=&
  e^{i[p(\xi_1)+p(\xi_2)+\pi]}|\xi_2,\epsilon_2;\xi_1,\epsilon_1;\pi\rangle.
  \nonumber
\end{eqnarray}
Since the 2-spinon momenta and energies
\begin{eqnarray}
P(\xi_1,\xi_2) &=& \bar{P}(\beta_1,\beta_2) \equiv p(\xi_1)+p(\xi_2),
\nonumber \\
E(\xi_1,\xi_2) &=& \bar{E}(\beta_1,\beta_2) \equiv e(\xi_1)+e(\xi_2),
\label{pe2}
\end{eqnarray}
are independent of the spin orientations $\epsilon_1,\epsilon_2=\pm 1$, all
2-spinon states at fixed $P$ will be at least fourfold degenerate.
In the isotropic limit, this degeneracy involves a singlet state $(S_T^z=S_T=0)$
and the three vectors with $S_T^z=0,\pm 1$ of a triplet state $(S_T=1)$.  

The four sets of 2-spinon excitations are readily identified in the framework of
the Bethe ansatz. In a finite system $(N<\infty)$, the singlet-triplet
degeneracy is removed, and for anisotropic coupling $(\Delta<-1)$, the triplet
levels are split up as well. The fourfold degeneracy emerges only asymptotically
for $N\to\infty$, and thus reflects the higher $U_q(sl_2)$ symmetry of the
infinite system, which is used in the algebraic analysis.
%
%
\section{Density of states}\label{secIII}
\begin{figure}[htb]
  \centerline{\epsfig{file=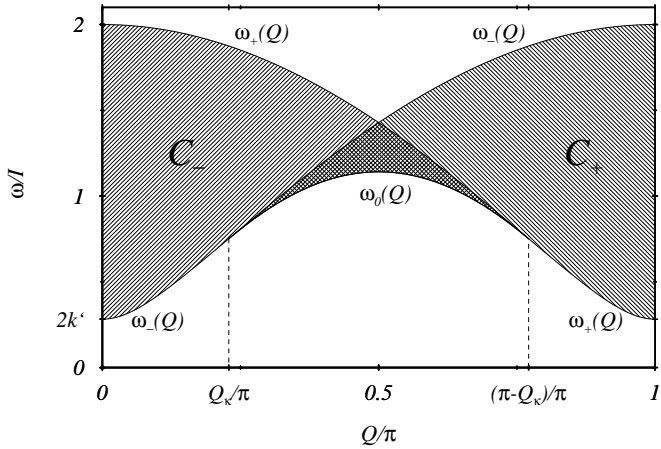,width=7.0cm,angle=-90}}
  \caption{Two-spinon excitation spectrum (\ref{MM21}) for $k=0.99$
    $(\Delta\simeq -2.305)$. It consists of two partly overlapping sheets
    $\mathcal{C}_-$ and $\mathcal{C}_+$. Sheet $\mathcal{C}_+$ lies between
    $\omega_0(Q)$ and $\omega_-(Q)$ in the range $Q_\kappa \leq Q \leq \pi -
    Q_\kappa$ and between $\omega_+(Q)$ and $\omega_-(Q)$ for $\pi-Q_\kappa
    \leq Q \leq \pi$.  Sheet $\mathcal{C}_-$ is obtained from $\mathcal{C}_+$ by
    reflection about the line $Q=\pi/2$. The interval $(0,\pi)$ represents one
    half of the extended Brillouin zone, in which one ground state vector,
    $|0\rangle$, is assigned wave number $Q=0$, and the other ground-state
    vector, $|\pi\rangle$, wave number $Q=\pi$.}
  \label{fig1}
\end{figure}

%
%
Here we consider any one of the four sets of 2-spinon excitations
(\ref{tev0pi2m}) with fixed spin orientations $\epsilon_1, \epsilon_2$ and
express their energies $E(\xi_1,\xi_2) = e_1(p_1)+e_2(p_2)$ in terms of the wave
number $Q=p_1+p_2$ $(0\leq Q < 2\pi)$ and the variable
$\lambda=\frac{1}{2}(p_1-p_2)$ $(-\pi/2 \leq \lambda < \pi/2)$:\cite{note6}
\begin{eqnarray}\label{MM21}
e_2(Q,\lambda) &\equiv& e_{1}(Q/2-\lambda) + e_{1}(Q/2+\lambda).
\end{eqnarray}
These states form a continuum in $(Q,\omega)$-space, which is depicted in
Fig.~\ref{fig1}. It consists of two partly overlapping sheets $\mathcal{C}_\pm$
with boundaries
\begin{mathletters}\label{MM25}
\begin{eqnarray}
  \omega_0(Q)    &=& \frac{2I}{1+\kappa}\sin Q, \label{MM25a} \\
  \omega_{\pm}(Q)&=& \frac{2I}{1+\kappa}\sqrt{1+\kappa^{2}\pm 2\kappa\cos Q},
\end{eqnarray}
\end{mathletters}
where  
\begin{equation}
  \label{cosqc}
  \kappa \equiv \cos Q_\kappa = \frac{1-k'}{1+k'}
\end{equation}
is the natural anisotropy parameter in most of the results presented here. The
excitation gap $\Delta E=2Ik'$ approaches zero exponentially in the isotropic
limit:\cite{dCG66}
\begin{equation}
 \Delta E \stackrel {\Delta\to -1}{\longrightarrow}
    4\pi J\exp\left(-\pi^2/\sqrt{-8(1+\Delta)}\right).
\end{equation}

\begin{figure}[htb]
  \centerline{\epsfig{file=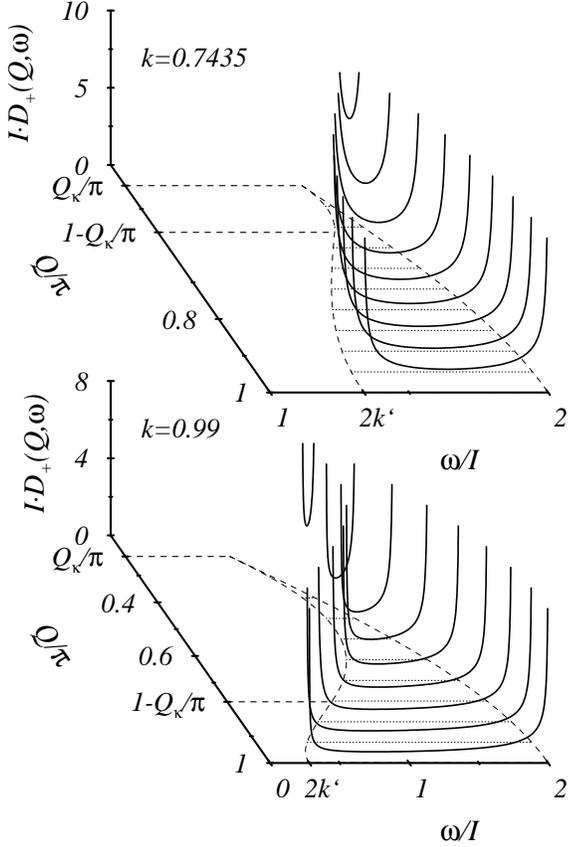,width=9cm,angle=0}}
  \caption{Normalized 2-spinon density of states $D_+(Q,\omega)$ as a function
    of frequency for wave numbers $Q_\kappa \leq Q \leq \pi$ and anisotropy
    parameter $k=0.7435$ ($\Delta\simeq -10$) and $k=0.99$ ($\Delta\simeq
    -2.305$).}
  \label{fig2}
\end{figure}

The 2-spinon density of states\cite{note3}
\begin{equation}\label{MM210}
  D(Q,\omega) \equiv \frac{1}{2}\int\limits_{-\pi/2}^{\pi/2} d\lambda\, 
                      \delta\biglb(\omega-e_2(Q,\lambda)\bigrb)
\end{equation}
was evaluated before in closed form:\cite{MM83}
\begin{mathletters}\label{D2pm}
\begin{eqnarray}
  D(Q,\omega)    &=& D_+(Q,\omega) + D_-(Q,\omega), \label{D2pma} \\
  D_\pm(Q,\omega)&=& n_\pm(Q,\omega)/d_\pm(Q,\omega),\label{D2pmb}
\end{eqnarray}
\end{mathletters}
for $(Q,\omega)\in\mathcal{C}_\pm$, respectively, and where
\begin{eqnarray*}
  \label{MM35and32}
  n_\pm(Q,\omega)  
  &=& 
  \frac{2\omega^2\!-\!(1\!+\!\kappa^2)\omega_0^2 
  \pm 2T\cos Q}{4\sin^2 Q},\\
  d_\pm(Q,\omega) 
  &=&
  \frac{T\left[(1\!+\!\kappa^2)\omega_0^2 \!-\! \omega^2(1+\cos^2 Q)
                    \mp 2 T \cos Q \right]^{1/2}}{2\sin^2 Q},
\end{eqnarray*}
\begin{equation}
  T(Q,\omega) = \sqrt{\omega^2-\kappa^2\omega_0^2} \sqrt{\omega^2-\omega_0^2}. 
  \label{TQw}
\end{equation}
With the auxiliary quantity
\begin{equation}\label{Wpm}
 W_\pm(Q,\omega) = 
    \sqrt{\frac{\omega_0^4}{\omega^4} \kappa^2 
      -\left(\frac{T}{\omega^2} \pm \cos Q\right)^2}
\end{equation}
the result (\ref{D2pmb}) can be written more compactly:
\begin{equation}\label{dpmqw}
 D_\pm(Q,\omega) = 
 \frac{\omega[\sin^2Q-W_\pm^2(Q,\omega)]}{2T(Q,\omega)W_\pm(Q,\omega)}.
\end{equation}
Note the reflection symmetry: $D_\pm(Q,\omega)=D_\mp(\pi-Q,\omega)$.  The
2-spinon density of state has square-root divergences all along the lower and
upper boundaries of each sheet. At the zone center, expression (\ref{dpmqw})
turns into
\begin{equation}
D_-(0,\omega)=\frac{\omega}{\sqrt{4I^2-\omega^2}\sqrt{\omega^2-4I^2k'^2}}.
\end{equation}
The function $D_+(Q,\omega)$ is plotted in Fig.~\ref{fig2} for two values of
anisotropy. 

%
%
\section{Matrix elements}\label{secIV}
%
%
The $m$-spinon eigenbasis provides a useful framework for the separate analysis
of the $m$-spinon contributions $(m=0,2,4,\ldots)$ to any zero-temperature
dynamical quantity of interest if a means of calculating the relevant matrix
elements can be found. Here we focus on the {\it 2-spinon} matrix elements of
the dynamic {\it spin} structure factor $S_{xx}(Q,\omega)$ at $T=0$.

With (\ref{comrel}) the 2-spinon part of (\ref{stfdef}) has the form
\begin{eqnarray}\label{stf2spj}
    S^{(2)}_{xx}(Q,\omega) &=& \frac{1}{8}
    \sum_{j,j'=0,1}\sum_{n=-\infty}^{+\infty} \sum_{\epsilon_1,\epsilon_2=\pm} 
   \; \int\limits_{-\infty}^{+\infty} dt e^{i(\omega t+Q n)}
\nonumber\\ \times
    \frac{1}{2}\oint \prod\limits_{i=1}^2 && \hspace{-5mm} 
                                            \frac{d\xi_i}{2\pi i\xi_i}
    {_{j'}\langle} 0|\sigma_n^x(t)|\xi_2\!,\!\epsilon_2;
                                 \xi_1\!,\!\epsilon_1\rangle_{j}
  {_j\langle}\xi_1\!,\!\epsilon_1;\xi_2\!,\!\epsilon_2|\sigma_0^x|0\rangle_{j'}.
\nonumber{} \\
\end{eqnarray}
The evaluation of this expression requires that we know all transition matrix
elements of the spin operators $\sigma_0^\pm = \frac{1}{2}(\sigma_0^x \pm
i\sigma_0^y)$ between the vacuum states $|0\rangle_j$ and the 2-spinon states
$|\xi_2,\epsilon_2;\xi_1,\epsilon_1\rangle_j$. All nonvanishing matrix elements
of this type turn out to be related to each other,
\begin{mathletters}\label{FF}
\begin{eqnarray}
  {_j\langle}0|\sigma_0^\pm|\xi_2,\mp;\xi_1,\mp\rangle_j 
&=&
{_{1-j}\langle}0|\sigma_0^\mp|\xi_2,\pm;\xi_1,\pm\rangle_{1-j}, \\
{_j{\langle }}\xi_1,\pm;\xi_2,\pm|\sigma_0^\pm|0\rangle_j 
&=& 
{_j\langle}0|\sigma_0^\pm|-q\xi_1,\mp;-q\xi_2,\mp\rangle_j,
\end{eqnarray}
\end{mathletters}
and can be expressed by a single function
\begin{equation}\label{Xjjm}
    X^j(\xi_2,\xi_1) = {_j\langle}0|\sigma_0^+|\xi_2,-;\xi_1,-\rangle_j ,
\end{equation}
which was determined by Jimbo and Miwa:\cite{JM95}
\begin{eqnarray}\label{Xjjm2}
    X^j(\xi_2,\xi_1) &=& \bar{X}^j(\beta_2,\beta_1) =
 \varrho^2\frac{\biglb(q^4;q^4\bigrb)^2}{\biglb(q^2;q^2\bigrb)^3}
\nonumber \\ && \hspace{-1cm} \times
  \frac{(-q\xi_1\xi_2)^{1-j}\xi_1
  \gamma(\xi_1^2/\xi_2^2) \theta_{q^8}(-\xi_1^{-2}\xi_2^{-2}q^{4j})}%
  {\theta_{q^4}(-\xi_1^{-2}q^{3})\theta_{q^4}(-\xi_2^{-2}q^{3})},
\end{eqnarray}
where
\begin{mathletters}\label{jmdef2}
\begin{eqnarray}
    \gamma(\xi)
&\equiv&
    \frac{\biglb(q^4\xi;q^4;q^4\bigrb)\biglb(\xi^{-1};q^4;q^4\bigrb)}%
         {\biglb(q^6\xi;q^4;q^4\bigrb)\biglb(q^2\xi^{-1};q^4;q^4\bigrb)}, \\
    \varrho
&\equiv&
  \bigrb(q^2;q^2\biglb)^2
  \frac{\biglb(q^4;q^4;q^4\bigrb)}{\biglb(q^6;q^4;q^4\bigrb)}, \\
    \biglb(x;y;z\bigrb)
&\equiv&
    \prod_{n,m=0}^{\infty}\left(1-xy^{n}z^{m}\right).
\end{eqnarray} 
\end{mathletters}

Carrying out the space-time Fourier transform and the sum over the spin
orientations in Eq.~(\ref{stf2spj}) yields
\begin{eqnarray}\label{stftrans}
  S^{(2)}_{xx}(Q,\omega) &=& \frac{1}{2}\left(\frac{\pi}{8K}\right)^2
    \int_{-2K}^{2K}\!\!d\beta_1\!\int_{-2K}^{2K}\!\!d\beta_2
    \delta[\omega-\bar{E}(\beta_1,\beta_2)] 
    \nonumber \\  && \hspace{-20mm} \times 
    \{\delta[ Q +\bar{P}(\beta_1,\beta_2)] 
    |\bar{X}^0(\beta_2,\beta_1)+\bar{X}^1(\beta_2,\beta_1)|^2 
    \nonumber \\  && \hspace{-20mm} +  
    \delta[ Q-\pi +\bar{P}(\beta_1,\beta_2)] 
    |\bar{X}^0(\beta_2,\beta_1)-\bar{X}^1(\beta_2,\beta_1)|^2\},
\end{eqnarray}
where we have also substituted (\ref{alphabeta}).

%
%
\section{Energy-momentum relations}\label{secV}
%
%
Performing the integrals over the spectral parameters in expression
(\ref{stftrans}) brings the 2-spinon dynamic structure factor into the form:
\begin{equation}\label{Sxxxi1xi2}
    S_{xx}^{(2)}(Q,\omega) = \frac{1}{2}
    \sum_{c=\pm}\sum_{\sigma=\pm} \frac{B_c^\sigma(Q,\omega)}{J_c(Q,\omega)},
\end{equation}
where the numerator,
\begin{eqnarray}\label{BQw}
    B_c^\sigma(Q,\omega)
&\equiv& 
   |\bar{X}^0(\beta^c_2,\beta^c_1)-\sigma \bar{X}^1(\beta^c_2,\beta^c_1)|^2,
\end{eqnarray}
is governed by the 2-spinon transition rates and the denominator,
\begin{eqnarray}\label{Jxi1xi2}    
    J_c(Q,\omega) &\equiv& 
    2\left(\frac{2K}{\pi}\right)^2\left|
    \frac{\partial\bar{E}}{\partial\beta_1}
    \frac{\partial\bar{P}}{\partial\beta_2} -
    \frac{\partial\bar{E}}{\partial\beta_2}
    \frac{\partial\bar{P}}{\partial\beta_1}\right|_{\beta_1^c\beta_2^c} 
    \nonumber \\
    && \hspace*{-1.5cm} = 2I\left(\frac{2kK}{\pi}\right)^2 
    \left|\rm{sn}\beta^c_1\rm{cn}\beta^c_1\rm{dn}\beta^c_2
     -\rm{sn}\beta^c_2\rm{cn}\beta^c_2\rm{dn}\beta^c_1 \right|,
\end{eqnarray}
by the 2-spinon density of states. In these expressions, the spectral parameters
now have fixed values $(\beta_1^c,\beta_2^c)$. These values are the solutions of
the 2-spinon energy-momentum relations arising from the two products of
$\delta$-functions in (\ref{stftrans}):
\begin{mathletters}\label{enemom}
\begin{eqnarray}\label{enemoma}
    \omega &=& \bar{E}(\beta_1,\beta_2),~~~~
   -Q = \bar{P}(\beta_1,\beta_2) \\
   \label{enemomb}
    \omega &=& \bar{E}(\beta_1,\beta_2),~~
    \pi-Q = \bar{P}(\beta_1,\beta_2),
\end{eqnarray}
\end{mathletters}
for $\sigma=\mp$, respectively.
Equations (\ref{enemom}) with (\ref{pe2}) and (\ref{pealpha}) are combined into
\begin{mathletters}\label{enemomt}
\begin{eqnarray}\label{enemomta}
\omega/I &=& \mathrm{dn}\beta_1 + \mathrm{dn}\beta_2, \\
\label{enemomtb}
-\sigma\sin Q &=& \mathrm{sn}\beta_1\mathrm{cn}\beta_2 +
\mathrm{cn}\beta_1\mathrm{sn}\beta_2,
\end{eqnarray}
\end{mathletters}
for the future analysis.

For fixed $\sigma$ and at a generic point $(Q,\omega)$ within the range of the
2-spinon continuum, there exists exactly one distinct solution per sheet
$\mathcal{C}_\pm$. Every such solution has multiplicity 8, accounted for by the
permutation symmetry $\beta_1 \leftrightarrow \beta_2$ of (\ref{enemomt})
[factor 2] and the periodicity of the elliptic functions [factor 4]. Now we use
addition theorems\cite{AS84} to convert (\ref{enemomt}) into
\begin{mathletters}\label{ep1ep2}
\begin{eqnarray} \label{e1e2bpm}
   \frac{\omega}{I} &=& 
         \frac{2 {\rm dn}\beta_+{\rm dn}\beta_-}
              {1-k^2{\rm sn}^2\beta_+{\rm sn}^2\beta_-}, 
\\ \label{p1p2bpm}
 -\sigma \sin Q &=& \frac{2{\rm sn}\beta_+{\rm cn}\beta_+{\rm dn}\beta_-}
              {1-k^2{\rm sn}^2\beta_+{\rm sn}^2\beta_-}
\end{eqnarray}
\end{mathletters}
with $\beta_\pm\equiv(\beta_1\pm\beta_2)/2$. From the ratio
\begin{equation}
  \label{sncndnp}
  -\frac{\sigma\sin Q}{\omega/I} = \frac{{\rm sn} \beta_+ {\rm cn} \beta_+}
                                {{\rm dn} \beta_+},
\end{equation}
we obtain
\begin{mathletters}\label{sncn}
  \begin{eqnarray}
    {\rm sn} \beta_+ &=& 
          -\sigma\sqrt{\frac{1}{2}
            \left(1 + \kappa\frac{\omega_0^{2}}{\omega^{2}} - 
                      \frac{T}{\omega^2}\right)}, \\
    {\rm cn} \beta_+ &=& 
          \sqrt{\frac{1}{2}
            \left(1 - \kappa\frac{\omega_0^{2}}{\omega^{2}} + 
                      \frac{T}{\omega^2}\right)}.
  \end{eqnarray}
\end{mathletters}
These solutions yield 
\begin{equation}\label{solution+}
  {\rm dn}\,\beta_+ = \sqrt{\frac{\omega^2-\kappa\omega_0^2 + T}%
  {\omega^2+\kappa\omega_0^2 + T}}
\end{equation}
and reduce (\ref{e1e2bpm}), effectively,
into a quadratic equation for $\mathrm{dn}\beta_-$ with $\sigma$-independent
solutions 
\begin{equation}\label{solution-}
 {\rm dn}\beta_- =  \frac{1\pm \cos Q}{\sin Q}
  \sqrt{\frac{\omega^2-\kappa\omega_0^2 + T}{\omega^2+\kappa\omega_0^2 - T}},
\end{equation}
where $\omega_0(Q)$ and $T(Q,\omega)$ are given in (\ref{MM25a}) and
(\ref{TQw}), respectively. Finally, a Landen transformation $(k\to\kappa)$
converts (\ref{solution+}) and (\ref{solution-}) into more explicit solutions in
terms of incomplete elliptic integrals,
\begin{mathletters}\label{landen}
\begin{eqnarray} \label{landenplus}
   \beta_+^c(Q,\omega) &=& -\sigma\frac{1+\kappa}{2} 
    F\left[\arcsin\frac{\omega_0}{\omega},\kappa \right],
      \\ \label{landenminus}
   \beta_-^c(Q,\omega) &=& \frac{1+\kappa}{2} 
    F\left[ \arcsin\left(
        \frac{2I\omega W_c}{\kappa(1+\kappa)\omega_0^2}
      \right),\kappa \right],
\end{eqnarray}
\end{mathletters}
where $W_c(Q,\omega)$ is given in (\ref{Wpm}), and the new label $c=\pm$
indicates that $(Q,\omega)\in \mathcal{C}_\pm$. 

\begin{figure}[htb] 
  \centerline{\epsfig{file=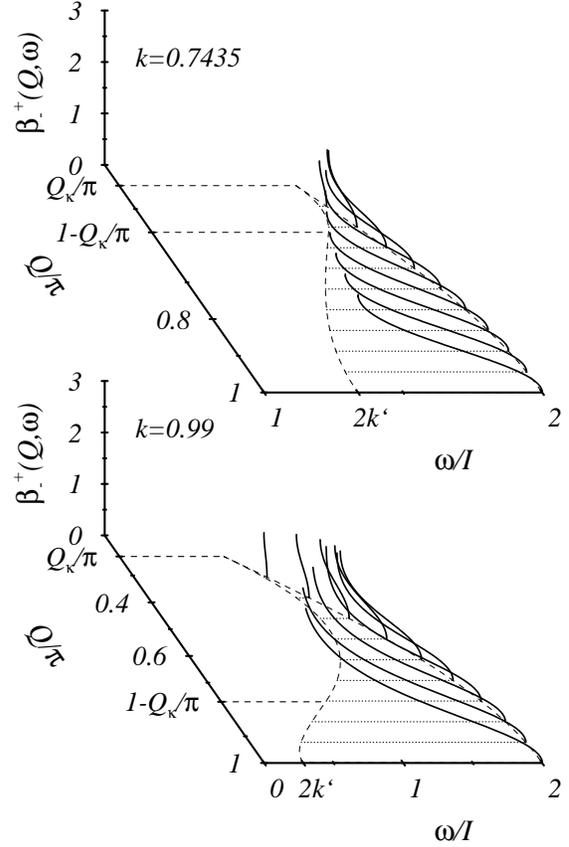,width=9cm,angle=0}}
  \caption{Explicit solution $\beta_-^+(Q,\omega)$
    of the 2-spinon energy-momentum equations as functions of frequency for wave
    numbers $Q_\kappa \leq Q \leq \pi$ and anisotropy parameters $k=0.7435$
    ($\Delta\simeq -10$) and $k=0.99$ $(\Delta\simeq-2.305)$.}
  \label{fig3}
\end{figure}

The function $\beta_-^+(Q,\omega) = \beta_-^-(\pi-Q,\omega)$, which alone enters
the final result, is plotted in Fig.~\ref{fig3} for two values of anisotropy. It
is finite along the lower boundary of $\mathcal{C}_+$, decreases monotonically
with increasing $\omega$ at fixed $Q$, and vanishes in a square-root cusp at the
upper boundary.  Note the different behavior along the portions $\omega_+(Q)$
and $\omega_0(Q)$ of the lower boundary of $\mathcal{C}_+$,
$\beta_-^+(Q,\omega_+) = K, \beta_-^+(Q,\omega_0) < K$, which will give rise to
different singularities in $S_{xx}^{(2)}(Q,\omega)$ in the two parts of the
spectral threshold.

%
%
\section{Dynamic structure factor}\label{secVI}
%
%
%
\subsection{Exact result for $S_{xx}^{(2)}(Q,\omega)$}\label{secVI.A}
%
The 2-spinon dynamic structure factor (\ref{Sxxxi1xi2}) for both sheets
$\mathcal{C}_\pm$ of the 2-spinon spectrum will now be evaluated as a product of
a density-of-state function $D_\pm(Q,\omega)$ and a transition-rate function
$M_\pm(Q,\omega)$, in generalization of the representation used in
Ref.~\onlinecite{KMB+97} for the $XXX$ model, and in analogy to the
representation used in Ref.~\onlinecite{MM83} for $S_{zz}(Q,\omega)$ of the $XY$
model:\cite{note7}
\begin{equation}\label{ScDcMc}
 S_{xx}^{(2)}(Q,\omega) = \frac{1}{2}\sum_{c=\pm}D_c(Q,\omega)M_{c}(Q,\omega).
\end{equation}
With the solutions (\ref{landen}) of the energy-momentum relations, the
numerator (\ref{BQw}) and the denominator (\ref{Jxi1xi2}) can be evaluated in
the form:
\begin{eqnarray}\label{BQwexp}
  B_c^\sigma(Q,\omega) = &&\left[\frac{2K(\kappa)}{\pi}\right]^2
  \frac{1+c\cos Q}{\omega_0^{2}}
  \frac{|A_-(\beta_-^c)|^2}{\vartheta_d^2(\beta_-^c)} \nonumber \\
  && \hspace{-15mm} \times \left[
    [\omega^2\!-\!\kappa\omega_0^2\!+\!T]\delta_{\sigma+}
   +\frac{1\!-\!\kappa}{1\!+\!\kappa}[\omega^2\!+\!\kappa\omega_0^2\!+\!T]
   \delta_{\sigma-}\right],
\end{eqnarray}
\begin{equation}\label{JQwexp}
  J_c(Q,\omega) = \left[\frac{4K(\kappa)}{\pi}\right]^2
                  \frac{\omega TW_c}{\omega_0^2}
\end{equation}
respectively, where
\begin{equation}
  \label{Abar}
    A_\pm(\beta) \equiv \!\exp\left(\! -\sum_{l=1}^\infty 
    \frac{\sinh^2[\gamma l(1\!-i\!\beta/K')]}{l\sinh(2l\gamma)\cosh(\gamma l)} 
    e^{\mp\gamma l} 
    \right)
\end{equation}
with $\gamma = \pi K'/K$, and $\vartheta_d(x)$ is a Neville theta
function.\cite{AS84} With the exact results (\ref{dpmqw}), (\ref{BQwexp}), and
(\ref{JQwexp}), the physically motivated factorization (\ref{ScDcMc}) of
$S_{xx}^{(2)}(Q,\omega) = \sum_{c=\pm}S_c(Q,\omega)$ can now be established:
\begin{equation}\label{dcqw}
 D_c(Q,\omega) \!=\! \left[\frac{4K(\kappa)}{\pi}\right]^2\!
  \frac{\omega^2[\sin^2Q\!-\!W_c^2(Q,\omega)]}{2\omega_0^{2}\,J_c(Q,\omega)}, 
\end{equation}
\begin{equation}\label{mcqw} 
 M_c(Q,\omega) \!=\! \sum_{\sigma=\pm 1} \!
 \left[\frac{\pi}{4K(\kappa)}\right]^2\!
  \frac{2\omega_0^{2}B_c^\sigma(Q,\omega)}%
  {\omega^2\,[\sin^2Q\!-\!W_c^2(Q,\omega)]}.
\end{equation}
The exact 2-spinon transition rate function thus obtained from (\ref{BQwexp})
substituted into (\ref{mcqw}) is
\begin{equation}\label{mcqwf}
 M_{c}(Q,\omega) = \frac{\omega^2-\kappa^{2}\omega_0^2+T}{(1+\kappa)\omega^2}
 \frac{1+c\cos Q}{\sin^2Q-W_c^2}
 \frac{\vartheta_A^2(\beta_-^c)}{\vartheta_d^2(\beta_-^c)},
\end{equation}
where $\vartheta_A^2(\beta) \equiv |A_-(\beta)|^2$ with (\ref{Abar}) is the
function 
\begin{equation}
  \vartheta_A^2(\beta) = \!\exp\left(\!-\sum_{l=1}^\infty 
      \frac{e^{\gamma l}}{l}
      \frac{\cosh(2\gamma l)\cos(t\gamma l)-1}%
      {\sinh(2l\gamma)\cosh(\gamma l)}
      \right),
\end{equation}
$\beta_-^c$ is given in (\ref{landenminus}), and $t\equiv 2\beta/K'$. The final
result for the exact 2-spinon dynamic structure factor reads:
\begin{eqnarray}\label{sxxfin}
 S_{xx}^{(2)}(Q,\omega) \!&=&\! \frac{\omega_0}{8I\omega}
    \left[ 1 \!+\! 
      \sqrt{\frac{\omega^{2}\!-\!\kappa^{2}\omega^{2}_0}%
        {\omega^{2}\!-\!\omega^{2}_0}}
    \right] \nonumber \\  && \times
    \sum_{c=\pm}  
    \frac{\vartheta_A^2(\beta_-^{c})}{\vartheta_d^2(\beta_-^c)} 
    \frac{|\tan(Q/2)|^{-c}}{W_c(Q,\omega)}
 \end{eqnarray}

%
\subsection{Line shapes and singularity structure}\label{secVI.B}
\begin{figure}[htb]
  \centerline{\epsfig{file=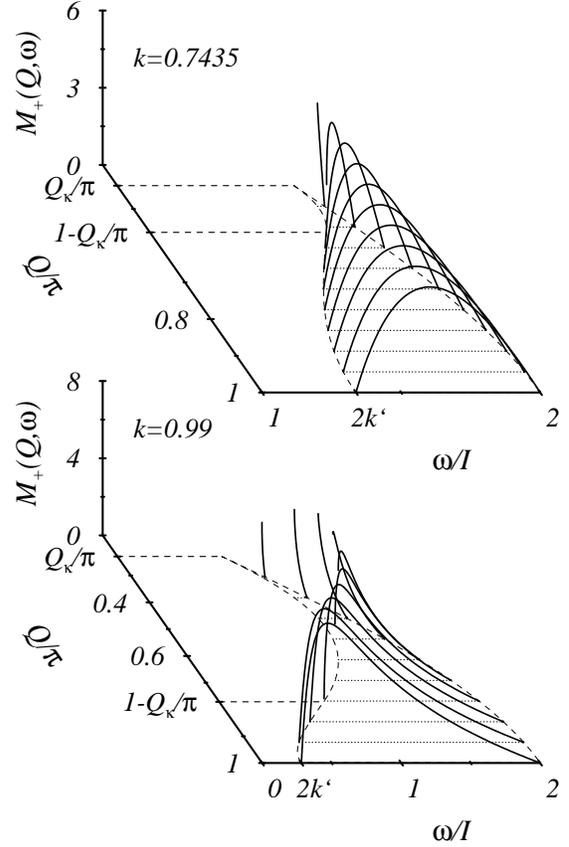,width=9cm,angle=0}}
  \caption{2-spinon transition rates $M_+(Q,\omega)$ as a function
    of frequency for wave numbers $Q_\kappa \leq Q \leq \pi$ and anisotropy
    parameters $k=0.7435$ ($\Delta\simeq -10$) and $k=0.99$
    $(\Delta\simeq-2.305$).}
  \label{fig4}
\end{figure}

\begin{figure*}[htb]
  \centerline{\epsfig{file=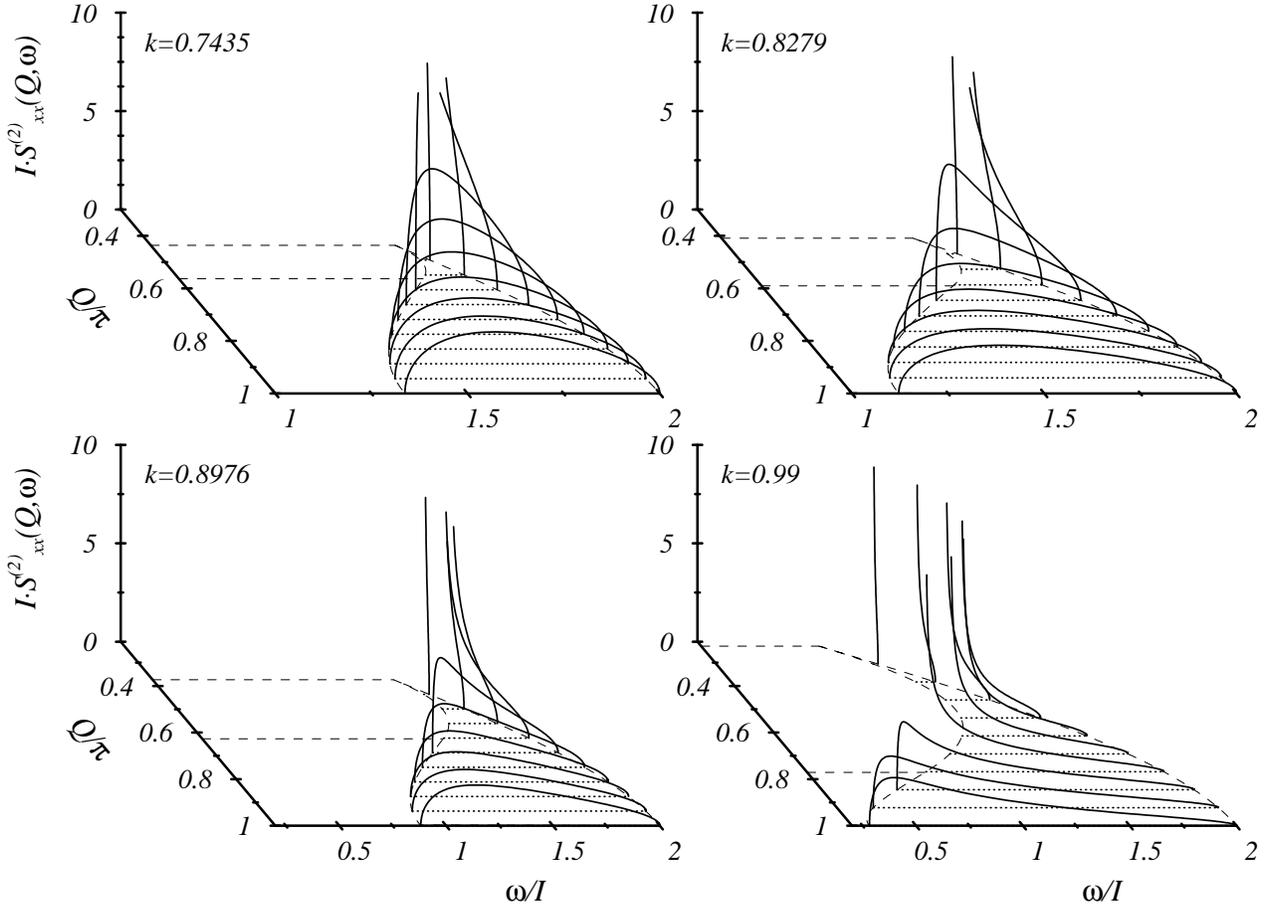,width=14.0cm,angle=-90}}
  \caption{2-spinon dynamic structure factor $S_{xx}^{(2)}(Q,\omega)$ for
    $(Q,\omega)\in\mathcal{C_+}$ as a function of frequency for wave numbers
    $Q_\kappa \leq Q \leq \pi$ and anisotropy parameters $k=0.7435$
    ($\Delta\simeq -10$), $k=0.8279$ ($\Delta\simeq -7$), $k=0.8976$
    ($\Delta\simeq -5$), and  $k=0.99$ ($\Delta\simeq-2.305$).} 
  \label{fig5}
\end{figure*}

%
The function $M_+(Q,\omega)$, which represents the 2-spinon transition rates for
$(Q,\omega)\in\mathcal{C_+}$, is plotted in Fig.~\ref{fig4} for two values of
anisotropy. The product of $M_+(Q,\omega)$ with the 2-spinon density of states
$D_+(Q,\omega)$ (already shown in Fig.~\ref{fig2} for the same two cases) yields
the spectral-weight distribution $S_+(Q,\omega)$ of the 2-spinon dynamic
structure factor for $(Q,\omega)\in\mathcal{C_+}$. This function is plotted in
Fig.~\ref{fig5} for four values of anisotropy, including the values chosen
in Figs.~\ref{fig2}-\ref{fig4}.

\widetext
\narrowtext

The transition rate function $M_+(Q,\omega)$ exhibits qualitatively different
properties in the two regimes $Q_\kappa \leq Q < \pi-Q_\kappa$ and $\pi-Q_\kappa
< Q \leq \pi$, where the spectral threshold is given by $\omega_0(Q)$ and
$\omega_+(Q)$, respectively (see Fig.~\ref{fig1}). In the first regime,
$M_+(Q,\omega)$ is nonzero at the lower boundary, decreases monotonically with
increasing $\omega$, and approaches zero linearly at the upper boundary. In the
second regime, by contrast, $M_+(Q,\omega)$ approaches zero linearly at both
boundaries and has a smooth maximum in between. The transition rates
for $(Q,\omega)\in\mathcal{C}_-$ are the exact mirror image: $M_-(\pi-Q,\omega)
= M_+(Q,\omega)$.

These properties of $M_\pm(Q,\omega)$ imply that the 2-spinon dynamic structure
factor $S_{xx}^{(2)}(Q,\omega)$ diverges all along the portion $\omega_0(Q)$ of
the spectral threshold and that the leading singularity there is the square-root
divergence of the 2-spinon density of states. Here the function $S_+(Q,\omega)$
decreases monotonically from infinity at the lower boundary to zero at the upper
boundary. Along the portion $\omega_+(Q)$ of the lower boundary and along the
entire upper boundary of $\mathcal{C}_+$, the linear behavior of the transition
rates removes the square-root divergence of the density of states in the product
and replaces it by a square-root cusp in the dynamic structure factor. Here the
spectral-weight distribution at fixed $Q$ has a smooth maximum between the band
edges. For strong anisotropy, the line shapes are broad and featureless. At
moderate to weak anisotropy, the line shapes are distinctly asymmetric with the
maximum positioned close to the spectral threshold.

The function $S_{xx}^{(2)}(Q,\omega)$, which is symmetric
about $Q=\pi/2$, is equal to one or the other of the two functions
$S_\pm(Q,\omega)$ except for $Q_\kappa \leq Q \leq \pi-Q_\kappa$. Here the two
sheets overlap, and the 2-spinon dynamic structure factor has two square-root
cusp singularities, one at the upper boundary of each sheet. The two upper
boundaries coincide only at $Q=\pi/2$.

%
\subsection{Isotropic limit}\label{secVI.C}
%
When we take the isotropic limit $\Delta\to-1^-$ in the results for the
2-spinon density of states, transition rates, and dynamic structure factor for
the purpose of linking up with results of calculations that were performed for
the Heisenberg model in the (non-degenerate) critical ground-state, we
must heed the fact that the size of the Brillouin zone changes from
$(-\pi/2,+\pi/2)$ to $(-\pi,+\pi)$ as the N\'eel long-range order in the ground
state of the infinite system vanishes. 

In practice, this means that we switch our perspective from considering both
sheets $\mathcal{C}_\pm$ of 2-spinon excitations over the range
$(-\pi/2,+\pi/2)$ to considering only the sheet $\mathcal{C}_+$ over the
extended range $(-\pi,+\pi)$. The term with $c=-$ in all expressions that
contain a sum $\sum_{c=\pm}$ is then be omitted. These contributions are
now accounted for in the term with $c=+$ over the extended range
of physically distinct wave numbers.

In the isotropic limit, the boundaries of $\mathcal{C}_+$ turn into the familiar
sine curves,
\begin{eqnarray}
  \omega_0(Q) &\to& \omega_L(Q) = \frac{\pi}{2}J\sin Q, \nonumber \\
  \omega_-(Q) &\to& \omega_U(Q) = \pi J \sin\frac{Q}{2},
\end{eqnarray}
and the 2-spinon density of states becomes
\begin{equation}
  D_+(Q,\omega) = \frac{1}{\sqrt{\omega_U^2(Q)-\omega^2}}.
\end{equation}

A major simplification occurs in the 2-spinon matrix elements,
\begin{equation}
X^1(\xi_2,\xi_1)~\to~-X^0(\xi_2,\xi_1),
\end{equation}  
which implies that all terms with $\sigma=-$ in (\ref{Sxxxi1xi2}) and
(\ref{BQwexp}) disappear in the isotropic limit. The results presented
previously for the isotropic case\cite{BCK96,KMB+97} can be recovered from the
more general result presented here, if we replace the spectral parameter $\beta$
in (\ref{alphabeta}) by the scaled spectral parameter $\alpha=-\beta\pi/2K'$,
and then take the limit $q\to-1^-$. In this limit, the auxiliary expressions
(\ref{jmdef2}) can be simplified considerably:
\begin{mathletters}\label{rhogamma}
\begin{eqnarray}
\varrho~&\to&~
  \frac{\biglb(q^4;q^4\bigrb)(1-q^4)^{1/4}}%
       {\biglb(q^2;q^2\bigrb)^2\Gamma(\frac{3}{4}) {\mathcal{A}}_+(i\pi/2)},
\\
  \gamma(\xi)~&\to&~
  \frac{\biglb(q^4;q^4\bigrb)(1-q^4)^{3/4}{\mathcal{A}}_-(\alpha)}%
       {\Gamma(\frac{1}{4}) {\mathcal{A}}_-(i\pi/2)},
\end{eqnarray}
\end{mathletters}
where
\begin{equation}
  \label{amb1b2}
  {\mathcal{A}}_\pm(\alpha) = 
  \exp\left( -\int\limits_0^\infty 
    \frac{dx}{x} \frac{\sinh^2[x(1+i\alpha/\pi)]}{\sinh 2x \cosh x} e^{\mp x} 
    \right)
\end{equation}
represents (\ref{Abar}) with $x=\gamma l$ in the limit $\gamma\to 0$.
The magnitude of the resulting 2-spinon matrix element becomes\cite{note4}
\begin{equation}
  \label{Xasymptotic}
  \left|X^0(\xi_2,\xi_1)\right|~ \to~ \frac{\pi}{2\gamma} 
  \left|\frac{{\cal A}_-(\alpha_1-\alpha_2)}{
    \sinh\left(\frac{i\pi}{4}+\frac{\alpha_1}{2}\right)
    \sinh\left(\frac{i\pi}{4}+\frac{\alpha_2}{2}\right)
     }\right|.
\end{equation}

Expression (\ref{mcqwf}) for $M_+(Q,\omega)$ reduces to
\begin{equation}
M_+(Q,\omega)=\frac{1}{2}e^{-I(t)},
\end{equation}
where 
\begin{equation}
   I(t) = \int_0^\infty dx \frac{\cosh 2x \cos xt -1}{x \sinh 2x \cosh x}e^x ,
\end{equation}
\begin{equation}
  \frac{\pi t}{4} = \frac{1}{2}(\alpha_1-\alpha_2) =  {\rm cosh}^{-1} 
    \sqrt{\frac{\omega_U^2(Q)-\omega_L^2(Q)}{\omega^2-\omega_L^2(Q)}}.
  \end{equation}
In Ref.~\onlinecite{KMB+97} this expression was evaluated, and its implications
for the 2-spinon dynamic structure factor\cite{note7}
\begin{equation}\label{ScDcMcI}
S_{xx}^{(2)}(Q,\omega) = M_+(Q,\omega)D_+(Q,\omega),~~~\Delta=-1
\end{equation}
were discussed in considerable detail.

%
\subsection{Ising limit}\label{secVI.D}
%
When we analyze the exact 2-spinon dynamic structure factor for very strong
anisotropy $(\Delta \to -\infty)$, a convenient expansion parameter about the
limiting Ising model is $\kappa$ as defined in (\ref{cosqc}).  Expressed in
terms of this parameter, the exchange anisotropy (\ref{Delta}) and the amplitude
of the 2-spinon dispersion (\ref{ampli}) become
\begin{equation}
|\Delta| \to \frac{2}{\kappa}\left(1+O(\kappa^{2})\right),~~
I \to \frac{1}{\kappa}\left(1+\kappa+O(\kappa^{2})\right).
\end{equation}
Here we set $J=1$. To lowest order in $\kappa$, the overlap region of the
two sheets $\mathcal{C}_\pm$ that make up the 2-spinon continuum (see
Fig.~\ref{fig1}) collapses to a single point at $Q=\pi/2$. The continuum
boundaries, as expressed in terms of the reduced frequency
\begin{equation}
\Omega \equiv \omega/2I - 1,
\end{equation}
are now described by the curves
\begin{equation}\label{ascbd}
\Omega_\pm(Q) \to \pm\kappa\cos Q.
\end{equation}
The regime $Q_\kappa \leq Q \leq \pi-Q_\kappa$ of the 2-spinon continuum is thus
squeezed to zero width at $Q=\pi/2$, where the 2-spinon bandwidth
is zero as well in lowest order.

The expansion of the 2-spinon dynamic structure factor can be carried out in the
final result (\ref{sxxfin}) by using
\begin{equation}
  W_c(Q,\omega) \to 
  \kappa\sin^{2}Q\sqrt{1-\left(\frac{\Omega}{\kappa\cos Q}\right)^{2}},
\end{equation}
\begin{equation}
  \vartheta_d(\beta_-^c) \to 1,~~~~
  \frac{\omega^2 - \kappa^2 \omega_0^2}{\omega^2 -\omega_0^2}
                           \to \frac{1}{\cos^{2}Q},
\end{equation}
 \begin{equation}
  \vartheta_A(\beta_-^{c}) \to 
  2\sin\left(2\beta_-^{c}\right) \simeq 
  \frac{4I\omega W_c}{(1+\kappa)\kappa\omega_0^{2}},
\end{equation}
yielding the closed-form expression
  \begin{equation}\label{eq:ising-dsf-zero}
    S_{xx}^{(2)}(Q,\omega) \to
    \frac{1}{2\cos^{2}Q}\sqrt{\cos^{2}Q-\left(\frac{\Omega}{\kappa}\right)^2},
  \end{equation}
which is identical (in lowest order) to the result
\begin{eqnarray}
  \label{eq:dsf-IS80}
  S_{xx}^{(IS)}(Q,\omega) &\to&
  \frac{\sqrt{4\cos^{2}Q-(\omega-|\Delta|)^{2}}}{4\cos^{2}Q}
  \nonumber  \\ &\times&
  \left(1-\frac{2}{|\Delta|}\left(\cos Q + \omega - |\Delta|\right)\right)
\end{eqnarray}
obtained by Ishimura and Shiba\cite{IS80} from a first-order perturbation
calculation about the Ising limit.

It is instructive to perform the expansion at an earlier stage of the
calculation, namely in (\ref{stftrans}), which then becomes
\begin{eqnarray}\label{sxxintas}
    S^{(2)}_{xx}(Q,\omega) &\to& \frac{1}{8}
    \int_{-\pi}^{\pi}d\beta_1 \int_{-\pi}^{\pi}d\beta_2
    \sin^{2}(2\beta_-) \delta(\omega-\bar{E}) \nonumber
    \\ &\times& 
    \left[\delta(Q +\bar{P}) +
      \delta(Q-\pi +\bar{P}) \right]
  \end{eqnarray} 
with
\begin{mathletters}
\begin{eqnarray}
  \bar{E}(\beta_1,\beta_2) &\to& \frac{2}{\kappa}
  \left[1+\kappa-\kappa(\sin^{2}\beta_1+\sin^{2}\beta_2)\right], \\
  \bar{P}(\beta_1,\beta_2) &\to& \pi - \beta_1 - \beta_2,
\end{eqnarray}
\end{mathletters}
and where we have used
\begin{mathletters}
\begin{eqnarray}
\bar{X}^0(\beta_1,\beta_2) &\to& 0, \\
|\bar{X}^1(\beta_1,\beta_2)|^2 &\to& 4\sin^2(2\beta_-).
\end{eqnarray}
\end{mathletters}
The asymptotic 2-spinon energy-momentum relations (\ref{enemomt}) are
\begin{mathletters}
\begin{eqnarray}
  -\omega+2+2/\kappa &=& 2(\sin^{2}\beta_1+\sin^{2}\beta_2), \\
  -\sigma\sin Q &=& \sin(\beta_1 + \beta_2),
\end{eqnarray}
\end{mathletters}
and the solutions
\begin{mathletters}
\begin{eqnarray}
\sin(2\beta_+^c) &=& -\sigma\sin Q, \\
\sin(2\beta_-^c) &=& \sqrt{1-\left(\frac{\Omega}{\kappa\cos Q}\right)^{2}},
\end{eqnarray}
\end{mathletters}
have multiplicity 8 for $(Q,\omega)$ within the boundaries (\ref{ascbd}) of the
asymptotic 2-spinon continuum. Performing the integrals in (\ref{sxxintas}) then
yields the asymptotic version of (\ref{Sxxxi1xi2}):
\begin{equation}\label{eq:ising-12}
  S^{(2)}_{xx}(Q,\omega) \to
  \sum_{\beta_1^{c},\beta_2^{c}}
  \frac{\sin^{2}(2\beta_-^{c})}{2|\sin(2\beta_-^{c})\cos(2\beta_+^{c})|},  
\end{equation}
from which (\ref{eq:ising-dsf-zero}) is recovered upon evaluation.

%
\subsection{Experiments}\label{secVI.E}
%
Two of the most intensively studied physical realizations of the 1D $s=1/2$
Heisenberg-Ising antiferromagnet are the quasi-1D magnetic compounds $\rm
CsCoCl_3$ and $\rm CsCoBr_3$. A very comprehensive set of spectroscopic data,
which probe diverse aspects of the low-temperature spin dynamics of these
materials, is now available from several inelastic neutron scattering
experiments performed over the course of 15 years at the
Brookhaven\cite{YHSS81}, Chalk River\cite{NBAB83a,NBAB83b},
Laue-Langevin\cite{BRRH85}, and Rutherford\cite{GTN95} Laboratories. This
impressive collection of data is complemented by a set of high-resolution
spectroscopic data for the long-wavelength spin dynamics obtained via Raman
scattering.\cite{LBW81}

The basis for the interpretation of all the experimental data that involve the
frequency range now known under the name 2-spinon continuum has been the
perturbation calculation about the Ising limit of (\ref{H}) carried out to first
order by Ishimura and Shiba\cite{IS80} (see also Ref.~\onlinecite{SS82}), which
yielded the explicit expression (\ref{eq:dsf-IS80}) for the $T=0$ dynamic
structure factor $S_{xx}(Q,\omega)$. Whereas this calculation does reproduce the
2-spinon continuum boundaries correctly to first order in the expansion
parameter, the reliability of its line-shape prediction -- a broad peak with the
maximum near the center of the band and steep drops near both boundaries -- has
remained very much in question.

The fact is that the line shapes observed in all experiments turned out to be
highly asymmetric with a high concentration of intensity near the spectral
threshold and a tail extending to the upper continuum boundary.

Various attempts have been made at reconciling the discrepancy between theory
and experiment by considering a second order perturbation calculation\cite{BC83}
for the pure $XXZ$ model, and by considering the impact of biaxial
anisotropy,\cite{BG96} next-nearest-neighbor coupling,\cite{MI91}
interchain coupling,\cite{Shib80} and exchange mixing,\cite{GTN95} all within
the framework of a first-order perturbation treatment.

In the range $-10 \lesssim \Delta \lesssim -7$ of anisotropies, which best
describes $\rm CsCoCl_3$ and $\rm CsCoBr_3$ according to some indicators, the
line-shape predictions obtained via perturbation calculation are in fair
agreement with the exact 2-spinon result for wave numbers near the zone center.
However, for wave numbers near the zone boundary, the exact 2-spinon
spectral-weight distribution is more asymmetric, consistent with the
experimental data. For $Q_\kappa < Q < \pi-Q_\kappa$, $S_{xx}^{(2)}(Q,\omega)$
even diverges at the spectral threshold and decreases monotonically toward the
upper continuum boundary. The conspicuous line-shape asymmetry found in the
experimental data near the zone center, which is not at all reproduced by the
perturbation calculation, does also exist in $S_{xx}^{(2)}(Q,\omega)$,
but only for weaker exchange anisotropy (see Fig.~\ref{fig5}).

%
\acknowledgments
%
This work was supported by the U.\ S. National Science Foundation, Grant
DMR-93-12252, and the Max Kade Foundation. The work at SUNYSB was supported by
NSF Grant PHY-93-09888.

%

\begin{references}
\bibitem{note5}
The quantum group $U_q(sl_2)$ is a deformation of the universal enveloping
  algebra $U(sl_2)$ of the Lie algebra $sl_2$: $[S_z,S_\pm] = \pm S_\pm,
  [S_+,S_-] = (q^{2S_z} - q^{-2S_z})/(q - q^{-1})$ with deformation parameter
  $q$. In the "classical" limit $q\to 1$, the familiar spin commutators are
  recovered.
\bibitem{note2} {The N\'eel long-range order breaks the translational symmetry
    of $H$ and thus reduces the Brillouin zone to $(-\pi/2,+\pi/2)$. In the
    isotropic limit $(\Delta\to -1)$ this symmetry is restored.}
\bibitem{note6}
The two sets of 2-spinon excitations which are relevant for
  $S_{xx}^{(2)}(Q,\omega)$ have $\epsilon_1=\epsilon_2=\pm1$.
\bibitem{note3}
From the Bethe ansatz we know that the density of 2-spinon states is uniform in
  $(Q,\lambda)$-space.
\bibitem{note7}
The factor $\frac{1}{2}$ present in (\ref{ScDcMc}) for $\Delta<-1$ and absent
  in (\ref{ScDcMcI}) for $\Delta=-1$ reflects the ground-state degeneracy
  ($T=0$ partition function) that enters the definition of the dynamic
  structure factor.
\end{references}


\bibliographystyle{prsty}
\end{document}